\documentclass[fleqn]{annalen}
\usepackage{psfig}
\usepackage{latexsym}
\begin{document}

\newcommand{\volume}{9}                  
\newcommand{\xyear}{2000}                
\newcommand{\issue}{3-5}                 
\newcommand{\recdate}{18 November 1999}  
\newcommand{\revdate}{dd.mm.yyyy}  
\newcommand{\revnum}{0}                  
\newcommand{\accdate}{18 February 2000}        
\newcommand{\coeditor}{ue}               
\newcommand{\firstpage}{288}             
\newcommand{\lastpage}{298}              
\setcounter{page}{\firstpage}     

\newcommand{\keywords}{cosmology, cosmic microwave background,
topological defects}
\newcommand{\PACS}{98.80.Cq, 98.70.Vc}

\newcommand{\shorttitle}
{A. Riazuelo et al., CMB anisotropies seeded by incoherent sources}
\title{Cosmic microwave background anisotropies seeded by incoherent sources}
\author{A.\ Riazuelo$^{1}$ and
        N.\ Deruelle$^{1,2,3}$}

\newcommand{\address}{
$^{1}$D\'epartement d'Astrophysique Relativiste et de Cosmologie, \\
UMR 8629 du CNRS, Observatoire de Paris, \\ 92195 Meudon, France \\
${}^2$Institut des Hautes \'Etudes Scientifiques, \\ 91140
Bures-sur-Yvette, France \\ $^{3}$DAMTP, University of Cambridge, \\
Silver Street, Cambridge, CB3 9EW, England}
\newcommand{\email}{\tt Alain.Riazuelo@obspm.fr} 
\maketitle

\begin{abstract}
The cosmic microwave background anisotropies produced by active seeds,
such as topological defects, have been computed recently for a variety
of models by a number of authors. In this paper we show how the
generic features of the anisotropies caused by active, incoherent,
seeds (that is the absence of acoustic peaks at small scales) can be
obtained semi-analytically, without entering into the model dependent
details of their formation, structure and evolution.
\end{abstract}


\newcommand{\ddd}{{\mathrm d}}
\newcommand{\Hconf}{{\mathcal H}}

\newcommand{\FTT}[1]{\hat{\bar{\bar{#1}}}}
\newcommand{\FVV}[1]{\hat{\bar{#1}}}
\newcommand{\FSS}[1]{\hat{#1}}
\newcommand{\TT}[1]{\bar{\bar{#1}}}
\newcommand{\VV}[1]{\bar{#1}}
\newcommand{\FFF}[1]{\hat{#1}}

\newcommand{\ETC}{etc}
\newcommand{\CF}{cf}
\newcommand{\IE}{i.e.}
\newcommand{\EG}{e.g.}
\newcommand{\ETAL}{{\it et al.}}
\newcommand{\EQ}[1]{eq.~(#1)}
\newcommand{\EQNS}[1]{eqns.~(#1)}
\newcommand{\FIG}[1]{fig.~{#1}}

\section{Introduction}

Two cosmological scenarios are currently in competition to explain the
large scale ($\theta> 7^\mathrm{o}$), fairly flat, spectrum of cosmic
microwave background (CMB) anisotropies which has been observed by the
COBE satellite~\cite{cobe}~: inflation (see, \EG~\cite{infl}), and
active seed models (see, \EG~\cite{td}).

Observations of the CMB anisotropy spectrum on smaller scales have
been performed by balloon and ground experiments~\cite{exp}, which
tend to indicate the presence of a peak at $\theta\simeq
1^\mathrm{o}$. This peak in the spectrum gives at present the lead to
the inflationary scenario.

One should however be careful not to exclude too hastily active seeds
at this stage. Indeed, to start with, the predictions from specific
topological defect models are not yet very robust, the reason being
that they act as a continuous, non-linear source of inhomogeneities,
and hence are difficult to model. Second, there exist simple models of
active seeds which ``mimic'' inflation and reproduce a peak in the
spectrum on the degree scale~\cite{mimic}. Third, including some
microphysics in the evolution of the defect network modifies the
``standard'' picture~\cite{rdp}. Even if such modellings are arguably
unrealistic, they show that the absence of a peak is not the seal of
all types of active seeds.

It is also important to be able to predict which active seeds produce
secondary peaks, like inflationary models, and which do not. As we
shall see, the absence of secondary peaks is the generic signature of
``incoherent'' seeds described within the ``stiff''
approximation. More then that the presence of a sharp peak on the
degree scale, the absence of secondary peaks, which will be probed by
the MAP and Planck satellite missions~\cite{map,plank} (and possibly
also by the Boomerang and Archeops~\cite{boomerang,archeops} balloon
experiments), will toll the bell of active seeds as the main
contributor to CMB anisotropies.

It is clear that the (semi-)analytic calculations of the seed
stress-energy tensor correlators pioneered by Durrer and
collaborators~\cite{durrer1} have the advantage, over heavy numerical
calculations within specific topological defect models, to test the
influence on the CMB anisotropies of each component of the seed
stress-energy tensor. This semi-analytic approach has already allowed
to grasp some generic features of the CMB anisotropies seeded by
active sources (see \EG~\cite{mimic,rdp,durrer1,turok2,dlu,udr}).

The object of this paper is to study, within this semi-analytic
approach, incoherent active seeds and show that, within the stiff
approximation, they all lead to CMB spectra exhibiting no secondary
peaks.

The paper is organised as follows~: in \S\ref{sec_spis}, we first
recall the some statistical properties of the seed stress-energy
tensor correlators at large wavelengths [see \EQ{\ref{pipi324}}], and
write some up-to-now unmentionned relations [see
\EQNS{\ref{vv12-vpi11}}]. We then give the new result that any active
source described by a test scalar field must obey these properties
[see \EQNS{\ref{pis}-\ref{pit}}].  In \S\ref{sec_sum}, we first recall
the definition and properties [see \EQ{\ref{pisvt_coh}}] of coherent
sources (\EG~\cite{dlu,udr}), and then explicitely construct for the first
time a generic (=incoherent) active source as a sum of coherent
ones. Finally, in \S\ref{sec_res}, we exhibit the generic behaviour of
the CMB anisotropies seeded by active sources.
  
\section{The statistical properties of incoherent sources}
\label{sec_spis}

\subsection{The 2-point correlators}

The stress-energy tensor of active sources is a small perturbation
added to the other cosmic fluids (this is the so-called ``stiff
approximation'' (see \EG~\cite{stiff})). We decompose its components
$\Theta_{\mu\nu}(x^i, \eta)$ into their scalar, vector and tensor
parts ($SVT$) as ($\mu, \nu = 0,1,2,3$, $\eta$ is the conformal time,
$x^i$ are $3$ cartesian coordinates; space is assumed to be flat for
simplicity)~:
\begin{eqnarray}
\label{t00}
\Theta_{00} & = & \rho^s, \\
\label{t0i}
\Theta_{0i} & = & -(\VV{v}_i^s+\partial_i v^s), \\
\label{tij}
\Theta_{ij} & = &
   \delta_{ij}P^s
 + \left(\partial_{ij}-\frac{1}{3}\delta_{ij}\Delta\right)\Pi^s
 + \partial_i\VV{\Pi}^s_j+\partial_j\VV{\Pi}^s_i
 + \TT{\Pi}^s_{ij} .
\end{eqnarray}
Barred spatial vectors are divergenceless; barred spatial tensors are 
traceless and divergenceless.

We work in Fourier space, the Fourier transform of any function
$f(x^i,\eta)$ being defined as~:
\begin{equation}
\hat f(k^i,\eta)=\int  e^{-{\rm i}k_ix^i}f(x^i,\eta) \ddd^3 x
\Longleftrightarrow  f(x^i,\eta)=\frac{1}{(2\pi)^3}\int
 e^{{\rm i}k_ix^i}\hat f(k^i,\eta) \ddd^3 k.
\end{equation}
From (\ref{t00}-\ref{tij}), we therefore have~:
\begin{eqnarray}
\label{x_SVT_s}
\left(k_i k_j - \frac{1}{3}\delta_{ij}k^2\right)\FFF{\Pi}^s
 & = & \left(   \frac{1}{2}P_{ij}P^{kl} + L^k_i L^l_j 
       \right) \FFF{\Theta}_{kl}, \\
\label{x_SVT_v}
2 k_{(i}\FVV{\Pi}_{j)}^s
 & = & 2 \left(P_{(i}^{(k} L_{j)}^{l)} \right) \FFF{\Theta}_{kl}, \\
\label{x_SVT_t}
\FTT{\Pi}_{ij}^s
 & = & \left(P_i^k P_j^l - \frac{1}{2}P_{ij}P^{kl}\right) \FFF{\Theta}_{kl} ,
\end{eqnarray}
and similar expressions for the other variables, with
\begin{equation}
\label{PL}
L_{ij} \equiv \delta_{ij} - P_{ij}
 \equiv \frac{k_i}{k}\frac{k_j}{k} \equiv \hat k_i \hat k_j.
\end{equation}

The ten components of $\Theta_{\mu\nu}(x^i,\eta)$ of the active source
are ten statistically spatially homogeneous and isotropic random
fields. The statistical properties of those ten random fields are
described by their unequal time two-point correlators
\begin{equation}
\left< \Theta_{\mu\nu}(x^i,\eta)
\Theta_{\rho\sigma}(x'^i,\eta')\right> \equiv
C_{\mu\nu\rho\sigma}(r^i,\eta,\eta'),
\end{equation}
where $\left< ...\right> $ means an ensemble average on a large
number of realisations, and where the correlator
$C_{\mu\nu\rho\sigma}$ is a tensor which depends only on $\eta, \eta'$
and $r^i\equiv x^i- x'^i$ because of the spatial homogeneity of the
distribution. The power spectra of the Fourier transforms are given
by~:
\begin{equation}
\label{corr_theta}
\left< \FFF{\Theta}^*_{\mu\nu}(k^i,\eta)
\FFF{\Theta}_{\rho\sigma}(k'^i,\eta')\right> =
\delta ( k^i- k'^i) \FFF{C}_{\mu\nu\rho\sigma}(k^i,\eta,\eta').
\end{equation}
The spatial isotropy of the distribution now forces the power spectra
to be of the form
\begin{eqnarray}
\label{p0000}
\FFF{C}_{0000} & = & A_0, \\
\FFF{C}_{000i} & = & ik_iB_1,\\
\FFF{C}_{00ij} & = & C_0\delta_{ij}+C_2k_ik_j, \\
\FFF{C}_{0i0j} & = & D_0\delta_{ij}+D_2k_ik_j, \\
\FFF{C}_{0ijk} & = & i\left[E_1k_i\delta_{jk} +\bar E_1(k_j\delta_{ik}
+k_k\delta_{ij})+E_3k_ik_jk_k\right], \\
\label{pijkl}
\FFF{C}_{ijkl} & = & F_0\delta_{ij}\delta_{kl}+\bar
F_0(\delta_{ik}\delta_{jl}+\delta_{il}\delta_{jk})+
F_2(k_ik_j\delta_{kl}+k_kk_l\delta_{ij})+ \\
\nonumber
 & & \bar F_2
(k_ik_k\delta_{jl}+k_ik_l\delta_{jk}+k_jk_l\delta_{ik}+k_jk_k\delta_{il})+
F_4k_ik_jk_kk_l,
\end{eqnarray}
where $A_0, B_1$ \ETC{} are 14 real functions of $\eta$, $\eta'$ and
the modulus $k$ of the spatial vector $k^i$.

We suppose that the active sources appeared at a definite time so that
the distribution must be, for causality reasons, completely
uncorrelated on scales larger than the particle horizon (we assume a
standard Big-Bang scenario).  Therefore, as stressed \EG{} by
Turok~\cite{turok2}, the unequal time correlators are strictly zero
outside the intersection of the past light-cones, that is~:
\begin{equation}
\label{corr_causal}
C_{\mu\nu\rho\sigma}(r^i,\eta,\eta') =0
\qquad\hbox{if}\qquad r>\eta+\eta'.
\end{equation}
Property (\ref{corr_causal}) translates in Fourier space into the fact
that the equal time power spectra are white noise on super-horizon
scales (that is for $k\eta\ll1$). Indeed, because the correlators
(\ref{corr_causal}) have compact supports, their Fourier transforms
are $C^\infty$ in $k^i$. Therefore causality forces the fourteen
functions $A_0$, $B_1$ \ETC{} to be $C^\infty$ in $k^2$. Moreover,
since within one horizon volume there are almost no sources, those
fourteen functions must tend to zero on small scales, that is for
$k\eta\gg1$. When $\Theta_{\mu\nu}$ is decomposed into its scalar,
vector and tensor components according to
\EQNS{\ref{t00}-\ref{tij}}, then
\EQNS{\ref{corr_theta}} and (\ref{p0000}-\ref{pijkl}) yield
\begin{eqnarray}
\label{vsvs}
\left<\FFF{V}^*_s\FFF{V}_s\right>     & = & D_0+k^2D_2, \\
\label{vvvv}
\left< \FVV{v}^{s*}_i\hat{\bar v}_j^s\right>   & = & 
  D_0 P_{ij},
\end{eqnarray}

\begin{eqnarray}
\label{vspis}
\left<\FFF{V}^*_s\FFF{\pi}_s\right> & = & -k(2\bar E_1+k^2E_3), \\
\label{vvpiv}
\left< \FVV{v}^{s*}_i\FVV{\pi}_j^s\right> & = & 
  -k\bar E_1P_{ij}, 
\end{eqnarray}

\begin{eqnarray}
\label{pispis}
\left<\FFF{\pi}^*_s\FFF{\pi}_s\right> & = & 
3\bar F_0+4k^2\bar F_2+k^4F_4, \\
\label{pivpiv}
\left<\FVV{\pi}^{s*}_i\FVV{\pi}_j^s\right>   & = & 
  (\bar F_0 + k^2 \bar F_2) P_{ij}, \\
\label{pitpit}
\left< \FTT{\Pi}^{s*}_{ij} \FTT{\Pi}_{kl}^s\right> & = & 
\bar F_0(P_{ik}P_{jl}+P_{il}P_{jk}-P_{ij}P_{kl}),
\end{eqnarray}
and similar expressions for the other correlators (see
\EG~\cite{udr}), where we have defined
$\left<\hat\rho^*_s(k^i,\eta)\hat\rho_s(k'^i,\eta')\right>
\equiv\delta(k^i-k'^i) \left<\hat\rho^*_s\hat\rho_s\right>$ \ETC, and
where we have introduced
\begin{equation}
\FFF{\pi}^s  \equiv  k^2 \FFF{\Pi}^s \quad,\quad
\FVV{\pi}_i^s  \equiv  k \FVV{\Pi}_i^s \quad,\quad 
V^s \equiv k v^s.
\end{equation}

An important property of these correlators is that since (for
causality reasons) $(k^2 \bar F_2)$ and $(k^4 F_4)$ are generically of
higher order in $k$ than $\bar F_0$, then, for small
$k$~\cite{turok2,udr}~:
\begin{equation}
\label{pipi324}
\frac{1}{3}\left<\FFF{\pi}^*_s\FFF{\pi}_s\right>
\simeq\frac{1}{2} \left< \FVV{\pi}^{s*}_i\FVV{\pi}^i_s\right>
\simeq \frac{1}{4}\left< \FTT{\Pi}^{s*}_{ij}\FTT{\Pi}^{ij}_s\right>.
\end{equation}
Similarly, when $(k^2 \bar D_2)$ is of higher order in $k$ than $D_0$,
and when $(k^2 \bar E_3)$ is of higher order in $k$ than $\bar E_1$,
one has~:
\begin{equation}
\label{vv12-vpi11}
\left<\hat v^*_s\hat v_s\right> \simeq 
\frac{1}{2}\left< \FVV{v}^{s*}_i \FVV{v}^i_s\right> \quad,\quad
\left<\FFF{v}^*_s\FFF{\pi}_s\right> \simeq 
\left< \FVV{v}^{s*}_i\FVV{\pi}^i_s\right>.
\end{equation}
 
These ratios are due to the geometric properties of the $SVT$
decomposition and the non-linear structure of the seed stress-energy
tensor. In order to give some insight about this phenomenon, we shall
present in the next paragraph the example of a scalar field.
 
\subsection{An example}

We consider a $N$-component real scalar field $\psi^A$ evolving in
an arbitrary potential $V$ according to the general Klein-Gordon
equations~:
\begin{equation}
\Box\psi^A + \frac{\partial V}{\partial \psi_A} = 0.
\end{equation}
This equation describes the evolution of a large class of topological
defects. The spatial traceless part of the stress-energy tensor
(denoted with a subscript $ST$) does not depend on $V$, and reads (see
also~\cite{kunz})~:
\begin{equation}
\FFF{\Theta}_{ij}^{ST}
 =   \widehat{\nabla_i \psi} * \widehat{\nabla_j \psi} 
   - \frac{1}{3}\delta_{ij} 
       \widehat{\nabla_r \psi} * \widehat{\nabla^r \psi} ,
\end{equation}
where a sum on the index $A$ is understood, and where $*$ stands for
the convolution product, so that
\begin{eqnarray}
\label{convol_1}
\FFF{\Theta}_{ij}^{ST} & = &
 - \int X_{ij} \psi(\vec p, \eta) \psi(\vec k- \vec p, \eta) 
   \ddd \vec p, \\
X_{ij} & = &   p_{(i} (k_{j)} - p_{j)})
             - \frac{1}{3}(\vec p . \vec k - p^2) \delta_{ij}.
\end{eqnarray}
We decompose $X_{ij}$ into its scalar vector and tensor parts, denoted
respectively by the superscripts $S$, $V$, $T$:
\begin{eqnarray}
\label{xis}
X_{ij}^S & = &
 p^2 \left(\frac{1}{3}\delta_{ij} - \hat k_i \hat k_j \right)
     \left(-\frac{1}{2} - \mu\frac{k}{p} + \frac{3}{2}\mu^2 \right), \\
\label{xiv}
X_{ij}^V & = & 
 p^2 \left(\frac{k}{p}-2\mu\right)
     \left(  \frac{\hat p_i \hat k_j + \hat k_i \hat p_j}{2} 
           - \mu \hat k_i \hat k_j\right), \\
\label{xit}
X_{ij}^T & = & p^2\left[
 \frac{1}{2}\delta_{ij}(1-\mu^2) -\frac{1}{2}\hat k_i \hat k_j(1+\mu^2)
-\hat p_i \hat p_j + \mu (\hat p_i \hat k_j + \hat k_i \hat p_j)
\right],
\end{eqnarray}
with the notations
\begin{equation}
\hat k_i \equiv k_i / k \quad,\quad 
\hat p_i \equiv p_i / p \quad,\quad 
\mu \equiv \hat p_i \hat k^i.
\end{equation}
Now, the correlators of the spatial traceless part of the
stress-energy tensor are~:
\begin{eqnarray}
\label{convol_2}
\left<\FFF{\Theta}_{ij}^{ST*}(\vec k, \eta)
      \FFF{\Theta}_{kl}^{ST}(\vec k', \eta')\right>
 & = & \int X_{ij}(\vec p, \vec k) X_{kl}(\vec p', \vec k') 
       \ddd \vec p \ddd \vec p' \\
\nonumber & \times &
  \left<\psi^*(\vec p, \eta)\psi^*(\vec k - \vec p, \eta)
  \psi(\vec p',\eta')\psi(\vec k'-\vec p',\eta')\right> .
\end{eqnarray}
Since the source distribution is homogeneous and isotropic, this
expression can be written under a form similar to
\EQNS{\ref{corr_theta}-\ref{pijkl}}~:
\begin{eqnarray}
\label{convol_3}
\left<\FFF{\Theta}_{ij}^{ST*}(\vec k, \eta)
      \FFF{\Theta}_{kl}^{ST}(\vec k', \eta')\right>
 & = & \delta(\vec k - \vec k') \\
\nonumber & \times &
\int  X_{ij}(\vec p, \vec k) X_{kl}(\vec p, \vec k) 
          Q(k, |\vec p-\vec k|, \eta, \eta') \ddd \vec p.
\end{eqnarray}
We see that the angular dependance arises only in the tensors
$X_{ij}$.  Moreover, in the limit $k \to 0$, one has
\begin{equation}
\int X_{ij}^S X^{ij}_S \ddd \mu \to \frac{4}{15} p^4 \quad,\quad
\int X_{ij}^V X^{ij}_V \ddd \mu \to \frac{2}{15} p^4 \quad,\quad
\int X_{ij}^T X^{ij}_T \ddd \mu \to \frac{8}{15} p^4.
\end{equation}
(The scalar, vector and tensor parts are always decoupled.) In
addition, we see that in the same limit $k \to 0$, all the correlators
(\ref{convol_3}) are proportional to the integral
\begin{equation}
\label{kto0}
I = \int p^6 Q(0, p, \eta, \eta') \ddd p ,
\end{equation}
(which is always non zero, being the average of a quartic term) so
that using \EQNS{\ref{x_SVT_s}-\ref{x_SVT_t},
\ref{convol_3}-\ref{kto0}}, one gets~:
\begin{eqnarray}
\label{pis}
\left<\FFF{\Pi}^s \FFF{\Pi}^s \right>
 = \frac{I\alpha}{k^4} \times 3 & \Longleftrightarrow &
\left<\FFF{\pi}^s \FFF{\pi}^s \right>
 =    I \alpha \times 3 , \\
\label{piv}
\left<\FVV{\Pi}_i^s \FVV{\Pi}^i_s \right>
 =  \frac{I\alpha}{k^2} \times 2 & \Longleftrightarrow &
\left<\FVV{\pi}_i^s \FVV{\pi}^i_s \right>
 =    I \alpha \times 2 , \\
\label{pit}
\left<\FTT{\Pi}_{ij}^s \FTT{\Pi}^{ij}_s \right>
 =  \frac{I\alpha}{k^0} \times 4,
\end{eqnarray}
where $\alpha$ is a numerical factor of order unity. It is easy to
show along similar lines that the correlators involving the velocities
obey \EQNS{\ref{vv12-vpi11}}.

Several conclusions arise from this simple, although generic,
model. The first is that, in the long wavelength limit, one finds, as
expected, that the scalar, vector and tensor correlators of the
anisotropic stress (\ref{pispis}-\ref{pitpit}) are in the ratio
$3:2:4$. The second is that this ratio is only an artefact of the
$SVT$ decomposition and of the fact that the seed stress-energy tensor
is quadratic in $\psi$~: it does not depend on the detailed dynamics
of $\psi$. However, due to the presence of the $k/p$ terms in the
$X_{ij}^{(a)}$, as well as the presence of $k$ in $Q$ [see
\EQNS{\ref{xis}-\ref{xit},\ref{convol_3}}], the angular dependance for
larger $k$ will no longer be in the same $3:2:4$ ratio~: the function
$\bar F_0$ defined in \EQ{\ref{pitpit}} is therefore proportional to
$I$, and the higher order terms in $k$ in $X_{ij}^{S,V}$ and $Q$
generate the expressions for $\bar F_2$ and $F_4$.

\section{Active sources as sums of coherent ones}
\label{sec_sum}

\subsection{Coherent sources}
\label{sec_coh}

By definition, a coherent source is such that the correlators
(\ref{p0000}-\ref{pijkl}) factorise~:
\begin{equation}
\label{corr_coh}
\FFF{C}_{\mu\nu\rho\sigma}(\eta, \eta', k^i) = 
\FFF{c}_{\mu\nu}(k^i, \eta) \FFF{c}_{\rho\sigma}(k^i, \eta') .
\end{equation}
Such a requirement implies, as shown in details in~\cite{udr},
that~:
\begin{eqnarray}
\label{f1f3}
&&\FFF{\rho}^s
 = \sqrt{\Hconf}     f_1(u, \eta) e(\vec k) \quad,\quad
\FFF{P}^s
 = \sqrt{\Hconf}     f_3(u, \eta) e(\vec k) ,\\
\label{f2f5}
&&\FFF{V}^s
 = \sqrt{\Hconf} u   f_2(u, \eta) e(\vec k) \quad,\quad
\FVV{V}_i^s
 = \sqrt{\Hconf} u   f_5(u, \eta) \VV{e}_i(\vec k) ,\\
\label{f4f6}
&&\FFF{\pi}^s
 = \sqrt{\Hconf} u^2 f_4(u, \eta) e(\vec k) \quad,\quad
\FVV{\pi}_i^s
 = \sqrt{\Hconf} u^2 f_6(u, \eta) \VV{e}_i(\vec k) ,\\
\label{f7}
&&\FTT{\Pi}_{ij}^s
 = \sqrt{\Hconf} u^2 f_7(u, \eta) \TT{e}_{ij}(\vec k),
\end{eqnarray}
where $\Hconf$ is the comoving Hubble parameter, where $u\equiv
k/\Hconf$. All the $f_i$ are arbitrary functions behaving as $u^0$
when $u \to 0$, and decaying to 0 when $u \to \infty$. When the $f_i$
do not depend on $\eta$, the sources have a scaling behaviour. The
$e$, $\VV{e}_i$, $\TT{e}_{ij}$ are independant complex random
variables which can be defined in such a way that
\begin{equation}
\label{ee}
\left< e(\vec k) e^*(\vec k')\right> = 
\left< \VV{e}^i(\vec k) \VV{e}_i^*(\vec k')\right> = 
\left< \TT{e}^{ij}(\vec k) \TT{e}_{ij}^*(\vec k')\right>
 = \delta(\vec k-\vec k').
\end{equation}
An immediate consequence of (\ref{f1f3}-\ref{ee}) is~:
\begin{equation}
\label{pisvt_coh}
\left< \FFF{\pi}^*_s\FFF{\pi}_s \right>\quad,\quad 
\left< \FVV{\pi}^{s*}_i\FVV{\pi}^i_s \right>\quad,\quad
\left< \FTT{\Pi}^{s*}_{ij}\FTT{\Pi}^{ij}_s \right>
 =  {\mathcal O}(k^4/\Hconf^3)
\end{equation}
and that, contrarily to the general case (\ref{pipi324}), the
anisotropic stress correlators of coherent sources are not in a
definite ratio for small $k$. The question to be asked is whether a
sum of coherent sources (with for example anisotropic stresses
correlators of order $k^4$) can lead to a generic incoherent source
(with anisotropic stresses correlators of order $k^0$).

\subsection{Coherent decomposition of a generic source}

As already stressed by several authors~\cite{kunz,pst,chm}, an
incoherent source can formally be decomposed into a sum a coherent
eigenmodes, that is its correlators can be written as~:
\begin{equation}
\FFF{C}_{\mu\nu\rho\sigma}(k, \eta,\eta')
 = \sum_{(i)} \lambda^{(i)} 
   \FFF{c}_{\mu\nu}^{(i)}(k,\eta) \FFF{c}_{\rho\sigma}^{(i)}(k,\eta').
\end{equation}
By coherent decomposition, we mean that the eigenmodes
$\FFF{c}_{\mu\nu}^{(i)}$ behave at low $k$ like the coherent sources
of \EQNS{\ref{f1f3}-\ref{f7}}. In a given topological model, these
eigenmodes are extracted from the $\FFF{C}_{\mu\nu\rho\sigma}(k,
\eta,\eta')$ given by the numerical simulation of the
network~\cite{kunz,pst,chm}.

In the semi-analytic approach adopted here, one can conversely
construct an active source by summing several functions $f_7$ (for the
tensorial part) behaving as $k^0$ but in such a way that their sum
behaves as $k^{-4}$ so as both \EQNS{\ref{pitpit}} and (\ref{f7}) are
satisfied. We show here a simple example of how this can be realized.

Let us consider a set of $N$ functions $f_n(k)$ obeying causality
and white noise constraints [\IE{} $f_n(k) \propto k^0$ when $k
\to 0$, and $f_n \to 0$ when $k \to \infty$]. For simplicity only,
we consider
\begin{equation}
f_n(k) = A_n Y(k_n -k),
\end{equation}
$A_n$ and $k_n$ being two sets of $N$ positive real numbers. We also
impose for simplicity that $\forall n, k_{n+1} < k_n$.  Let us
consider then
\begin{equation}
F_N = \sum_n f_n.
\end{equation}
It is easy to see that
\begin{equation}
F_N(k_0) = \sum_n f_n(k_0) = \sum_{n=0}^{n_0} A_n \quad\hbox{with}\quad 
n_0 \equiv \max\{n|k_n > k_0\} .
\end{equation}
We then consider a function $F$ which is decreasing on ${\bf R}_+$ and
tends towards $\infty$ around $0$ (for example, $F(k) =
k^{-\alpha}$). It is straightforward to see that by choosing
\begin{equation}
\label{dec_a_nn}
A_n = 1/\sqrt{N} \quad,\quad k_n = F^{-1}(n/\sqrt{N}),
\end{equation}
the set $F_N$ converges towards $F$. We have therefore explicitely
built a function behaving as $1/k^\alpha$ with an (infinite) sum of
functions behaving individually as $k^0$. Coming back to the
correlators, the cost is that the causality constraints fade away as
we increase the number of functions in the sum since $x_n \to 0$ when
$N \to \infty$.  However, the causality constraints need not be
satisfied by the coherent eigenmodes, only the sum requires causality
to be satisfied.

The fact that we have taken a Heaviside function instead of a more
regular function is not important. A set of gaussians of various
widths would give the same result at the leading order in $k$. The
fact that we use a finite truncated sum is unimportant as well, as
long as the eigenmodes we neglect influence scales much bigger than
today's Hubble radius only.

\section{Results and conclusion}
\label{sec_res}

\begin{figure}
\begin{center}
\centerline{\psfig{file=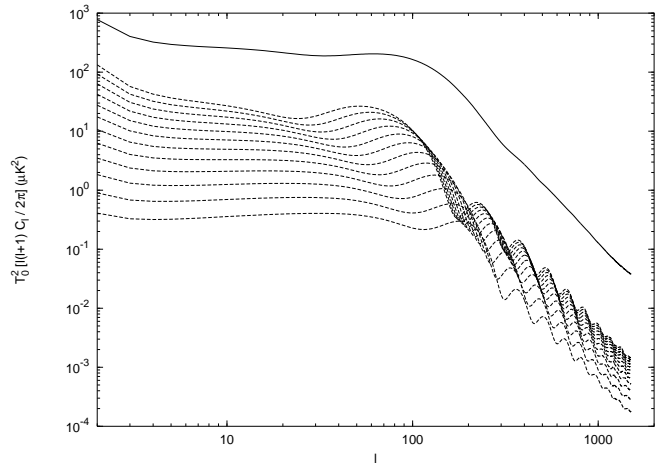,width=3.5in,angle=-90}}
\end{center}
\caption{Analytic construction of the tensorial contribution of incoherent
active sources to the CMB anisotropies. The incoherent source (solid
line) is a sum of 12 coherent eigenmodes (dashed lines). The
correlators of the coherent eigenmodes behave as $k^4$, whereas the
correlator of the incoherent source behaves as $k^0$. Vertical units
are arbitrary.}
\label{fig_inc}
\end{figure}

We considered here tensor modes only, and we computed their
contribution to the CMB anisotropies by means of a Boltzmann code
developped by one of us (A.R., see~\cite{cmb} for the detailed
equations). We took a sum of coherent and scaling modes such that
\begin{equation}
\FTT{\pi}_{ij}^{s,(n)}
 = \sqrt{\Hconf} u^2 \sqrt{A_n} \exp(-k_n^2 u^2/2) \TT{e}_{ij},
\end{equation}
with $A_n$ and $k_n$ given by \EQNS{\ref{dec_a_nn}}, such that
\begin{equation}
\left<\FTT{\pi}_{ij}^s \FTT{\pi}^{ij}_s\right> = 
\sum_n \left<\FTT{\pi}_{ij}^{s,(n)} \FTT{\pi}^{ij}_{s,(n)}\right>
 \propto k^0.
\end{equation}
The results are shown on \FIG{\ref{fig_inc}}. We see that the
sum converges, and that some of the modes have roughly the same
amplitude in the region of the Doppler peaks, but not the same phase~:
the Doppler peak structure is washed out.

The disappearance of Doppler peaks comes from the fact that although
each coherent eigenmode gives an oscillatory contribution to the CMB
anisotropy, their sum does not. Indeed, as $n$ increases, the
extension in $u$ of $\FTT{\pi}_{ij}^{s,(n)}$ decreases. This implies
that the metric is perturbed for shorter and shorter duration hence on
larger and larger scales.

The fact that one has to consider a sum of several eigenmodes comes
from \EQ{\ref{convol_3}}, which is a convolution product, hence is not
factorisable as a product of functions
$\FFF{c}(k,\eta)\FFF{c}(k,\eta')$. However, it is possible that one
eigenmode dominates the sum, in which case the Doppler peak structure
can partially remain, as seems to be the case in the large $N$
model~\cite{dkm}.

Finally, the convolution product appearing in \EQ{\ref{convol_2}}
comes from the fact that any stress-energy tensor component is (a
least) \emph{quadratic} with respect to the perturbation (\IE{} the
field $\psi$), therefore any correlator is at least quartic with
respect to the field. In the most optimistic case, by using Wick's
theorem (this is possible when one studies the large $N$ model,
see~\cite{dkm}), this quartic contribution can be reduced to a
quadratic one, but even in this case, the remaining formula
\EQ{\ref{convol_2}} is not factorizable as a product of two
quantities.

On the contrary, in inflationary scenarios (and, more generally, in
any scenario producing an initial power spectrum of fluctuations), the
cosmic fluids (baryons, neutrinos, \ETC) are described as coherent
sources. Indeed their correlators are only quadratic in the
perturbations, because there exists a contribution of the fluids to
the background. For example, one has (forgetting here the metric
perturbation)~:
\begin{eqnarray}
\FFF{\delta T}_{00} & \propto & \rho(\eta) \FFF{\delta}(k, \eta), \\ 
\FFF{\delta T}_{ij} & \propto & P(\eta) \FFF{\pi}_{ij}(k, \eta),
\end{eqnarray}
and so on, and solving the equations of evolution, one will find that
the perturbations are \emph{linear} with respect to the initial
conditions (as in the case of the large $N$ model, but here, the
stress energy-tensor is also linear with respect to the
perturbations)~:
\begin{equation}
\FFF{\delta}(\vec k, \eta) =  
 \sum_{(a)} F_{(a)}(k, \eta) \FFF{X}_{(a)}(\vec k, \eta^*),
\end{equation}
where $X_{(a)}$ stands for all the $\delta$, $V$, $\pi$ of the
different species which take part in the initial conditions (at
$\eta=\eta^*$). Then if one calculates the correlators of the
stress-energy tensor components, one obtains a quadratic dependance
with respect to the perturbations, which can be cancelled by using the
correlators of the initial conditions. For example~:
\begin{eqnarray}
\nonumber
 & & \left<\FFF{\delta T}_{00}^*(\vec k,\eta), 
           \FFF{\delta T}_{00}(\vec k', \eta')\right> \\
\nonumber
 & \propto &  \sum_{(a),(b)} \rho(\eta) \rho(\eta') 
              F_{(a)}(k, \eta) F_{(b)}(k', \eta') 
              \left<\FFF{X}_{(a)}^*(\vec k, \eta^*)
                    \FFF{X}_{(b)}(\vec k', \eta^*)\right> \\
 & \propto & \sum_{(a),(b)} \rho(\eta) \rho(\eta') 
              F_{(a)}(k, \eta) F_{(b)}(k', \eta') 
              \delta(\vec k - \vec k') \FFF{C}_{(ab)}(k).
\end{eqnarray}
Moreover, all the perturbations generally are supposed (but this is
not always the case, see~\cite{rl}) to depend on only one random
variable so that the sum in the last expression is reduced to one
component.  We also see that one will not get any convolution product
of the form $F_{(a)}(p, \eta) F_{(b)}(|\vec k - \vec p|, \eta')$, and
therefore the correlators will factorise.

In conclusion, topological defects are incoherent because their
stress-energy tensor is quadratic with respect to the field, whereas
for standard fluids (\IE{} fluid which have a contribution to the
background) it is only linear. Therefore, the fact that defects are
incoherent is ultimately linked with the stiff approximation.

\vspace*{0.25cm} \baselineskip=10pt{\small \noindent It is a pleasure
to thank Martin Kunz and Jean-Philippe Uzan for helping us to clarify
some points about the ${\cal O}(N)$ model, and Ruth Durrer for
numerous, fruitful, enjoyable discussions.}

\end{document}